\documentclass[sigconf, 10pt]{acmart}
\renewcommand\footnotetextcopyrightpermission[1]{}
\graphicspath{{./figures/}}
\usepackage{array}
\usepackage{hyperref}
\usepackage{url}
\usepackage{flushend}

\begin{document}
\title[Detecting Compressed Cleartext Traffic from Consumer Internet of Things Devices]{Detecting Compressed Cleartext Traffic from\\Consumer Internet of Things Devices}

\author{Daniel Hahn}
\affiliation{%
  \institution{Princeton University}
  \city{Princeton}
  \state{New Jersey}
}
\email{danielph@princeton.edu}

\author{Noah Apthorpe}
\affiliation{%
  \institution{Princeton University}
  \city{Princeton}
  \state{New Jersey}
}
\email{apthorpe@cs.princeton.edu}

\author{Nick Feamster}
\affiliation{%
  \institution{Princeton University}
  \city{Princeton}
  \state{New Jersey}
}
\email{feamster@cs.princeton.edu}

\renewcommand{\shortauthors}{D. Hahn et al.}

\begin{abstract}
Data encryption is the primary method of protecting the privacy of consumer device Internet communications from network observers.  
The ability to automatically detect unencrypted data in network traffic is therefore an essential tool for auditing Internet-connected devices. 
Existing methods identify network packets containing cleartext but cannot differentiate packets containing encrypted data from packets containing compressed unencrypted data, which can be easily recovered by reversing the compression algorithm. 
This makes it difficult for consumer protection advocates to identify devices that risk user privacy by sending sensitive data in a compressed unencrypted format. 
Here, we present the first technique to automatically distinguish encrypted from compressed unencrypted network transmissions on a per-packet basis.  
We apply three machine learning models and achieve a maximum 66.9\% accuracy with a convolutional neural network trained on raw packet data. 
This result is a baseline for this previously unstudied machine learning problem, which we hope will motivate further attention and accuracy improvements. 
To facilitate continuing research on this topic, we have made our training and test datasets available to the public. 
\end{abstract}

% CXML


\settopmatter{printfolios=true}

\maketitle

\section{Introduction}

Many government agencies, industry standards groups, and academic researchers have identified best practices for transmitting private consumer information over the public Internet. A cornerstone of these practices is the use of transport layer encryption to prevent eavesdropping network observers from obtaining sensitive consumer data.  Unfortunately, the rush to market for new IoT or other smart devices can sometimes cause developers to cut corners, neglecting encryption best practices and placing consumer data at risk. 

While this is a concern for all networked devices, the Federal Trade Commission (FTC) has noted that ``companies entering the Internet of Things (IoT) market may not have experience in dealing with security issues,'' making them particularly likely to overlook network data encryption \cite{FTC:iot}. 
Lack of encryption for consumer IoT devices is especially concerning because many have always-on environmental sensors that constantly record user behaviors, often inside homes or other private spaces, and stream these recordings to the cloud \cite{apthorpeNoCastle}. 
Additionally, IoT devices often have limited user interfaces, making it difficult for users to determine the extent to which data is being collected, transmitted, or protected.

It is therefore paramount that researchers and consumer advocacy groups have effective tools to audit Internet-connected devices and identify privacy concerns.  
The results of such audits are often the only information consumers have by which to choose IoT devices that provide desired privacy protections. 
Negative audit results motivate manufacturers to improve their products, while positive results help consumers feel more confident that their devices do not exposing sensitive data.

Previous research has identified numerous violations of best practice guidelines for network data encryption by companies producing IoT devices, including cases where sensitive consumer medical information was sent in cleartext \cite{Wood:clear}.
While not all cleartext data transmissions pose major privacy risks, the FTC has identified cases where the observation of unencrypted data can be used to infer extremely specific user behavior or for various forms of identity theft \cite{FTC:iot}.

While tools exist to flag cleartext network transmissions from IoT devices \cite{Wood:clear}, they fail to detect communications that have been compressed rather than encrypted, since they rely on assumptions about entropy and normal data distributions which are similar between compressed and encrypted data. 
Yet, compressed unencrypted data is no more secure than unmodified cleartext; a simple reversal of the compression algorithm is enough to recover the information, requiring no knowledge of secret keys. 

This project frames the differentiation of unencrypted compressed packets from encrypted packets as a binary classification machine learning task, with the goal of developing a method for the automated detection of all unencrypted data transmissions regardless of encoding.  
The method operates on a per-packet basis rather than on reconstructed files in order to facilitate device auditing by ISPs and other entities with incomplete or sparse packet captures. While this methodology intends to provide a tool for the analysis of IoT consumer device traffic, its intended use case is not in the consumer space. Instead, this project aims to provide researchers with a faster and more integrated pipeline with which to flag packets that may represent consumer privacy violations.

We generated training and test datasets by collecting a diverse set of 200 files, compressing some with the most popular compression algorithms, encrypting the rest with common encryption algorithms, transferring all files through a loopback TCP connection, and capturing the resultant packets.  

We trained three machine learning models to distinguish packets containing data from compressed versus encrypted files using only packet contents and no TCP header information. 
A 5-layer convolutional neural network achieved the highest accuracy, correctly identifying 66.9\% of of packets with compressed unencrypted data. 
While this accuracy leaves plenty of room for future improvement, it indicates that the task is indeed possible despite the similarity of character distributions between encrypted and compressed data. 
This result should therefore serve as a baseline for this previously unstudied machine learning task, which we hope will attract more attention and motivate follow-up work to improve classification accuracy. 
To facilitate further research, we have made our labeled training and test datasets available to the public at \url{https://github.com/danielph312/comp_detect_csv}.

\section{Related Work}

\subsection{Cleartext Detection}
Wood et al. describe a method to identify cleartext information in network
traffic using a Chi-Square test, where the observed frequency of each byte is compared to
its expected value from a uniform distribution \cite{Wood:clear}. If a packet contains encrypted
or compressed data, its expected entropy will be very high, and the Chi-Square of its byte distribution value will be low,
since we expect a relatively uniform distribution of bytes in encrypted data. If a packet contains
cleartext, however, its expected entropy will be lower, since natural language cleartext would not be
expected to conform to a uniform distribution. This method yields a high success rate in identifying
cleartext packets with virtually no false-positives. Unfortunately, this method cannot differentiate
between semantically secure encrypted data, which will be completely random by definition, and compressed data, from
which cleartext can be recovered with a fast brute-force decompression approach \cite{conte2014techniques}. A different approach is needed to distinguish these two cases. 

\subsection{Distinguishing Compressed vs.$\!$ Encrypted Files}
There are a number of existing general solutions to differentiate encrypted files from compressed files. We will highlight two previous approaches for detecting compressed files. However, they are either ineffective or inefficient for classifying network packets, especially if all traffic from a device is not recorded (either due to network vantage point or intentionally sparse packet captures). 

\subsubsection{Brute Force}
US Patent 8799671 B2 \cite{conte2014techniques} describes a largely brute-force attack wherein files that are identified as encoded undergo attempted decompression with a variety of popular compression algorithms and are assumed to be encrypted upon failing to decompress with any of these algorithms \cite{conte2014techniques}. 
While this approach is effective for files, it is not applicable for detecting compressed unencrypted network traffic on a per-packet basis. The nature of compression algorithms means that an entire file is needed for decompression; a single packet containing a portion of a compressed file is insufficient. Reconstructing complete files from network captures, however, requires a prohibitive degree of human intervention for the analysis of large network sessions and may not be possible if the packet capture is sparse.  The tools available for file reconstruction neglect TCP transfers in lieu of explicit file-oriented transfer protocols such as FTP and HTTP. While TCP file carvers do exist, they are often inaccurate, and rely on individual file markers differentiating the beginnings and ends of files in a data stream  \cite{Deck:packets}. Additionally, the reliance on these markers makes them completely ineffective at carving encrypted data, which appears completely random, or at carving homebrew file formats that contain unrecognized file markers, making this approach not particularly useful for the problem at hand. Since recovering full files for an entire network transfer is prohibitively difficult, US Patent 8799671 B2 is not directly useful for flagging compressed packets. However, it is a practical method by which to confirm the compression status of a recovered file, if a reconstruction can be completed \cite{Deck:packets}. 

If file reconstruction using the methods identified in Stephen Deck's paper \cite{Deck:packets} is impossible given a series of packets flagged as compressed, then the file would have to be reconstructed manually, increasing human labor overhead. Therefore, automatic detection of packets containing compressed unencrypted data requires a very high accuracy rate, as such packet streams may take a nontrivial amount of effort to reconstruct and evaluate for privacy concerns.

\subsubsection{Compression Fingerprinting}
\begin{figure}[t]
\centering
\includegraphics[width=0.47\textwidth]{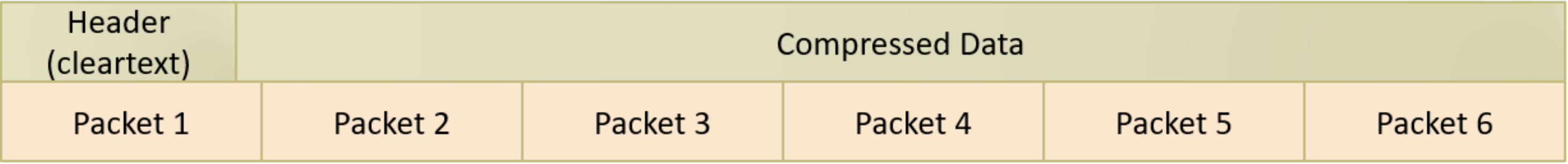}
\caption{Dividing a compressed file into packets relegates the most distinctive characteristics of the file to a small percentage of the overall packets transmitted.}
\label{fig:headercom}
\end{figure}

Paras Malhotra describes qualitative strategies for differentiating encrypted and compressed data based on particular fingerprints left by gzip and bzip, two popular compression algorithms \cite{Mal:streams}. 
However, the data fingerprints in question typically occur at the beginning of a given file, but files may be split into thousands of packets in a typical TCP stream \cite{pckt}, of which only one will contain the crucial fingerprint (Figure~\ref{fig:headercom}). This makes Malhotra's approach impractical for per-packet classification. Therefore, it would be advantageous to identify packets that may be compressed, instead of files, such that a file reconstruction can then be performed to confirm the encoding type and examine the contained data without the need to attempt to perform a file reconstruction on each potentially compressed TCP transfer.

\subsection{Note on TLS}
By best practices, all encrypted data should be sent across a secure TLS channel. 
TLS application data packets are easily identified by their headers with no further classification techniques required. 
However, TLS requires a certificate, which takes time and effort to implement. 
Instead, some developers may choose to use homebrew encryption (since high-quality public cryptography libraries are free and easily available), masquerade compression as encryption, ignore data security altogether, or employ some other less-than-secure solution \cite{EU:iot}. Strategies like this are
of particular concern in the IoT consumer space, where software development costs and development time are crucial in a rapidly evolving device marketplace \cite{FTC:iot}.
We therefore focus on non-TLS TCP connections in an attempt to distinguish compressed data from data encrypted using an alternative encryption algorithm.  

\section{Data Collection}

\begin{figure}[t]
\centering
\includegraphics[width=0.47\textwidth]{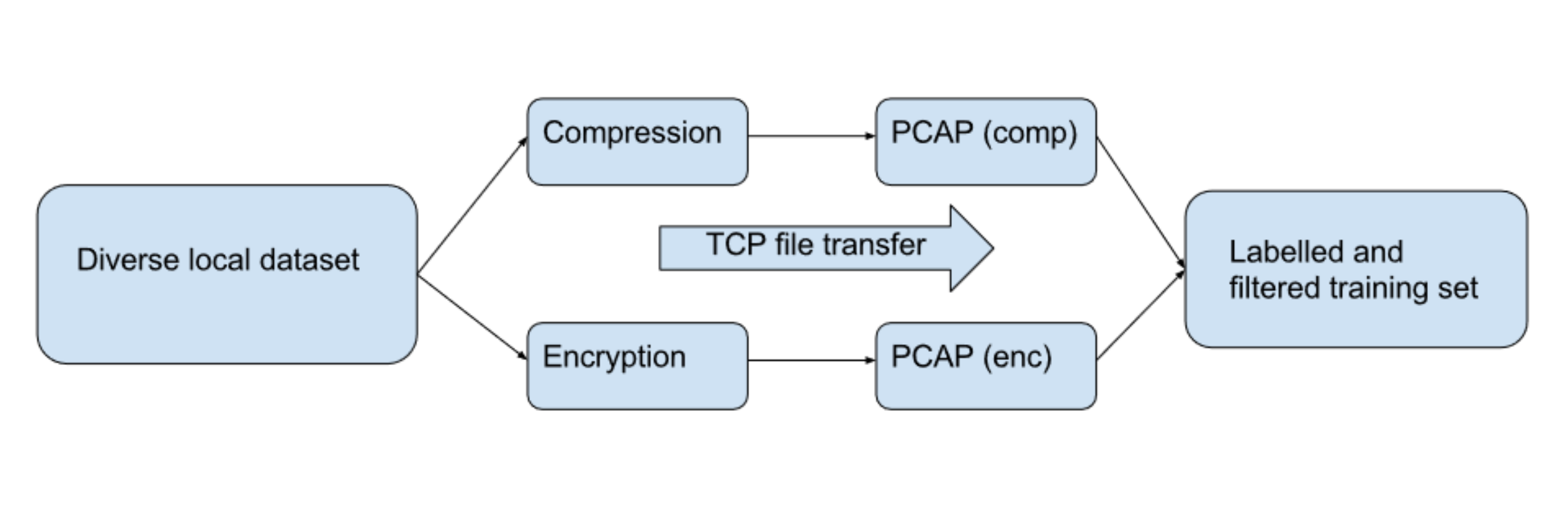}
\vspace{-15pt}
\caption{A simplified overview of the data collection process. Data is compressed or encrypted, transmitted over localhost, and a packet capture is generated from the resultant network transfer.}
\label{fig:pipeline}
\end{figure}

The supervised machine learning models we tested require labeled training data. Given the lack of publicly available packet captures with pre-labeled compressed and encrypted packets, we generated our own dataset with the following procedure. (Figure~\ref{fig:pipeline}). 
We first collected a set of 200 unique files ranging between 1 KB and 2478 KB. The files included a range of filetypes, including PDFs, text files, pictures including PNG, GIFs and JPGs, HTML documents, binaries, and executables. We split these files into equally sized compression and encryption groups, attempting to maintain a similar distribution of file sizes between groups. The compression group was further divided into five sets of 20 files each. Each set was compressed using one of the five of the most commonly used compression algorithms: bzip2, 7zip, zip, rar, or tar.gz. The encryption group was further divided into 2 sets of 33 files and one of 34.  Each set was encrypted using AES, Blowfish, and RC5, respectively. 

We then transferred all files over a TCP loopback connection to localhost in random order within the compressed/encrypted groups, with uniformly randomized packet sizes of 500 to 1500 bytes per file transfer to
approximate typical packet size for this transmission protocol \cite{pckt}. 
We used the application RawCap to record packet capture (PCAP) files from localhost loopback, providing PCAPs containing solely packets of either compressed or encrypted data with a range of packet sizes.  We then used the Python library dpkt to extract packet contents and tag each with a classification bit (1 for compressed, 0 for encrypted). The final result was a CSV file containing rows of bytes terminating in a classification bit, allowing for direct use in training and testing. While this process provided a large and fairly normal dataset for model training, additional preprocessing was required to account for the input requirements of each machine learning model.

This dataset is available at \url{https://github.com/danielph312/comp_detect_csv}. The dataset was trimmed slightly by 385 compressed packets to provide an equal sample size of compressed and encrypted packets, giving us 8398 examples of each in the 16796 row CSV file. Since the files were shuffled prior to network transfer, the truncation of a few hundred packets from the compressed data does not disproportionately affect the representation of any individual algorithm. The provided dataset contains packets fixed at 1024 bytes to allow for use with neural nets, as discussed further in this paper. This data collection strategy attempts to emulate the general types of files that may be transmitted by IoT applications over a TCP connection by sampling a wide range of both user data and application data.

\section{Implementation}

\begin{figure}[t]
\centering
\includegraphics[width=0.47\textwidth]{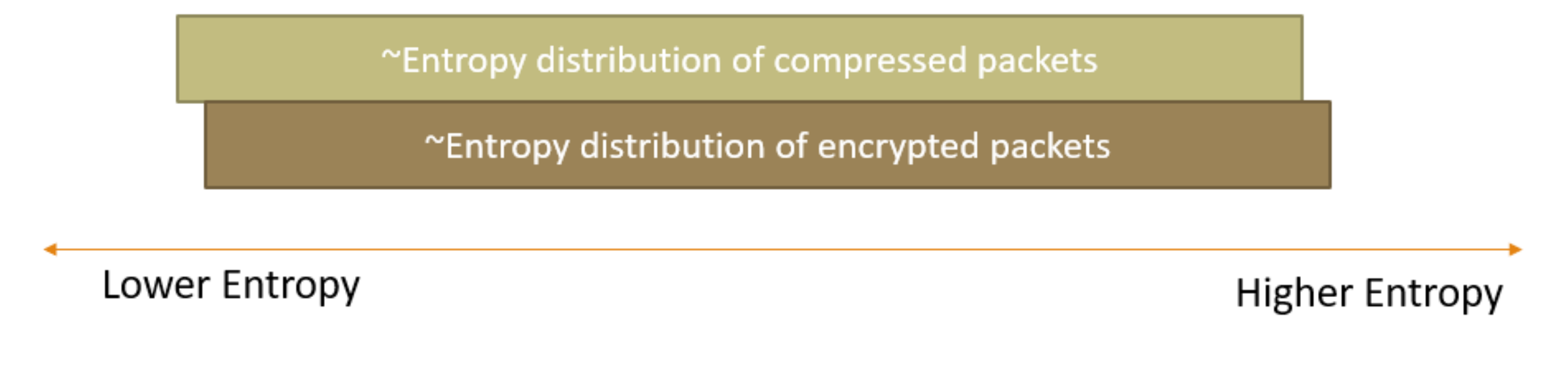}
\vspace{-10pt}
\caption{A generalized representation of the differences in entropy between compressed and encrypted packets. Compressed packets are 99.5\% as entropic as encrypted packets, while each group has a large standard deviation of entropy across packets.}
\label{fig:entropy}
\end{figure}

As noted by Malhotra \cite{Mal:streams}, many compression algorithms have a fingerprint of some sort. While the clearest indication of compression is in file headers, there also exists ``a small difference in the [overall] entropy levels of an encrypted stream and a compressed stream'', and compressed files inherit some degree of non-randomness from their corresponding uncompressed
files \cite{Mal:streams}. 
This suggests that automated differentiation of compressed and encrypted traffic should be possible. 
Without relying on assumptions about individual algorithms, this difference in
entropy is undetectable to either a Chi-Square test or a Shannon Entropy Test.
Analysis of our dataset shows that the difference in average entropy between a large set of generated compressed packet contents and a corresponding set of encrypted packet contents is much smaller than the standard deviation of entropy within each category (Figure~\ref{fig:entropy}); we observed an average Shannon entropy of 1.149 in compressed files and 1.154 in encrypted files and a standard deviation of 0.451 in compressed files and 0.453 in encrypted files. This minimal entropy difference is likely attributable to compression headers (which are low entropy), further reducing the entropy distinction across the majority of compressed packets.
This means that thresholding based on entropy is an ineffective classification method, confirming the claims of D. Wood et al. regarding the possibility of classifying compressed packets~\cite{Wood:clear}.

Our approach instead focuses on machine learning methods to discover usable nonlinear decision boundaries and high level features by which to preform compression/encryption classification, avoiding the pitfalls of simple threshholding. We implemented three such classification techniques: k-nearest neighbors, a feed-forward neural network, and a 1D convolutional neural network.

\subsection{k-Nearest Neighbors}

\begin{figure}[t]
\centering
\includegraphics[width=0.47\textwidth]{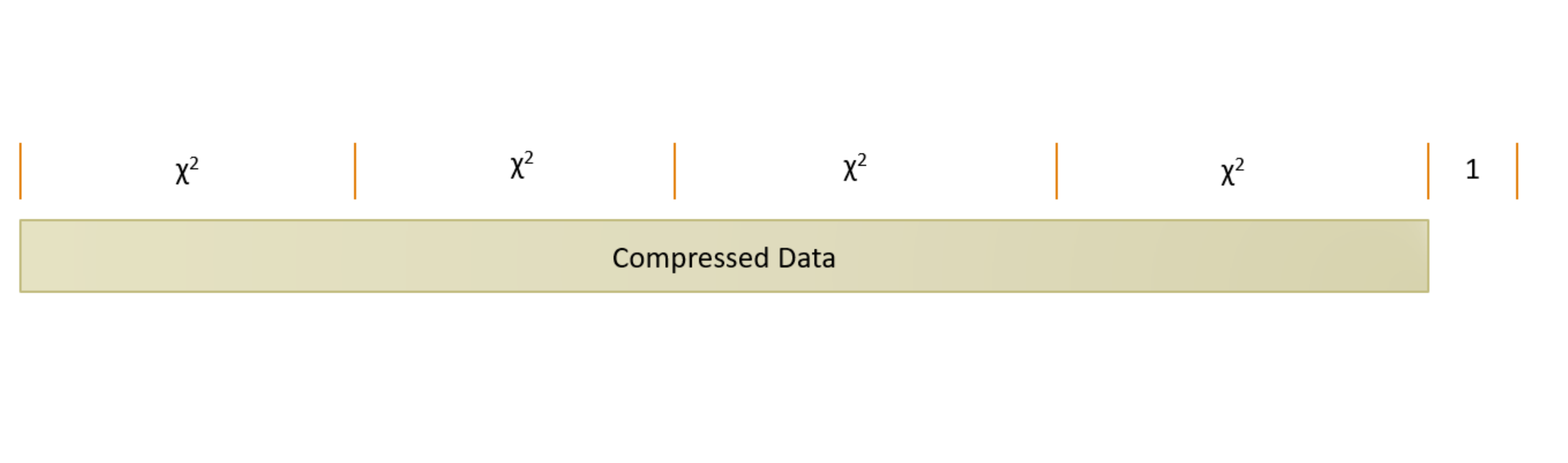}
\vspace{-25pt}
\caption{A representation of the data format used in k-NN classification. Chi-Square analysis is performed on each quadrant of the file, and then concatenated to generate a dataset with rows containing 4 Chi-Square columns and one classification column.}
\label{fig:chi}
\end{figure}

A k-nearest neighbors (k-NN) model was tested as a simple baseline requiring minimal training effort.
Since k-NN compares a candidate to its neighbors in feature space, high-level features are not extracted by the model in the manner they would be in a neural network.
Therefore, manual identification of features is required. 
The features employed by our k-NN model are semi-local Chi-Square analyses of each packet. 
Individual packets in the training and testing sets are divided into four equal chunks of data. 
A Chi-Square distribution analysis comparing each chunk to a uniform distribution of possible byte values converts each packet into a length 4 feature vector of $\chi^2$ statistics (Figure~\ref{fig:chi}). 
Using four features allows for a model with basic nonlinear decision boundaries, potentially avoiding the previously discussed difficulty posed by entropy thresholding.
There could theoretically be more effective features, but the process of discovering them is challenging and relies on educated guessing. We also tested features based on localized Shannon entropy, nonlocalized Shannon entropy, and nonlocalized Chi-Square statistics, but all were less effective than localized Chi-Square statistics as feature representations.

We trained a k-NN classifier using the Python SciKit-learn library \cite{scikit-learn}.
To fine-tune model parameters, we used a validation set of 3000 compressed and 3000 encrypted packets. 
For the purposes of packet classification, the most effective parameters were $k = 9$ with a normalized weighting, meaning that each query packet was be compared to its 9 nearest neighbors, with weighting directly proportional to distance in feature space. 

This methodology is easily employable in real-world network capture analysis, since Chi-Square feature extraction can be preformed on packets of any length. Reducing the number of features to a static four-way localized entropy analysis allows this model to function on any packets with minimal preprocessing and without padding or other means of feature dimension normalization, which would be required by models that use raw packet data and require an input of fixed dimensionality. 

\subsection{Feed-forward Neural Network}

A feed-forward neural network is distinct from k-NN in that it requires no explicit features aside from the raw packet payload data. 
Despite this, feed-forward neural networks expect an input with a fixed number of degrees. 
This presents issues in the real world, where packet sizes can be wildly variable.
A potential solution would be to pick a number of bytes such that the large majority of non-cleartext packets are greater in size than this threshold. 
Packets exceeding this size could be truncated, while packets under this length could be padded or discarded. 
Fortunately, in the development and evaluation of such a network, we have the luxury of standardizing packet sizes to determine the general efficacy of such an approach. For the purposes of this network, we re-ran the data generation pipeline with packet sizes fixed at 1024 bytes. This allowed the raw data of a huge volume of packets to serve as our neural network training set.

The neural network was built using the Keras library \cite{chollet2015keras} with a TensorFlow-GPU backend. Keras enables the construction of feed-forward networks with explicitly defined layers. We experimented with a number of network architectures, but settled on a network with 3 ReLU layers with 256, 128, and 128 neurons, followed by a 1 neuron sigmoid layer. The kernel initialization was normal for every layer, and the network was trained using Adam optimization \cite{DBLP:journals/corr/KingmaB14} on binary cross-entropy loss with a batch size of 5.

\subsection{Convolutional Neural Network}

Convolutional neural networks (CNNs) are currently the de facto standard for difficult classification problems using raw data. 
While typically used for image classification, CNNs are effective across data types and problem domains~\cite{simard2003best}. 
The local connectivity of convolutional neural networks makes them ideal for identifying regional features \cite{1606.00298}, making them an promising choice for classifying packets based on minute compression fingerprints identified by Malhotra \cite{simard2003best} \cite{Mal:streams}. 

We implemented a 1D CNN using Keras with a TensorFlow-GPU backend.
The CNN, like the linear neural network, took raw packet payloads as input represented as 1D vectors of length 1024. 
The first convolutional layer had 64 filters with kernel size 16 and ReLU activation followed by a standard Keras MaxPooling1D layer. 
The second convolutional layer had 32 filters with kernel size 16 and ReLU activation followed by a MaxPoooling1D layer. 
The final layer was a fully-connected sigmoid layer with one neuron, allowing the complete model to output binary classifications.

The model was trained on 80 epochs with a batch size of 50 using the adam optimizer and a binary crossentropy loss function.

\section{Evaluation}
\label{sec:evaluation}
\begin{figure}[t]
\centering
\begin{tabular}{ll}
\hline
\textbf{Classifier} & \textbf{Accuracy} \\
\hline
k-Nearest Neighbors & 60.0 \\
Feed Forward Neural Net & 54.1 \\ 
Convolutional Neural Net & 66.9 \\
\hline
\end{tabular}
\caption{Accuracy rates of various approaches when evaluated on test sets containing equal numbers of positive/negative examples.}
\label{fig:acc}
\end{figure}

There is no exact threshold for success for any of the described methods. Compression/encryption classification has never been attempted as a machine learning problem, so there are no existing baselines against which to compare our results. However, a successful model would be 
\begin{enumerate}
  \item Accurate enough to enable recovery of compressed files from packets without needing to attempt much brute-force decompression or file carving on misclassified packets
    \item Fast enough to preform reasonably efficient classification for device auditing on a customized router or programmable switch
\end{enumerate}
The following sections discuss the accuracy results of each tested model (Figure~\ref{fig:acc}).

\subsection{k-Nearest Neighbors}
k-NN performance was tested using 10-fold cross validation on the training set. Each fold used 3000 compressed packets and 3000 encrypted packets for training and 700 and 700 packets for testing, respectively.  The k-NN achieved an average  accuracy of 60.0\% across all folds. These results are clearly statistically significant; a random model would converge on a 50\% overall accuracy over many trials. Additionally, the model was more eager to falsely classify compressed packets as encrypted than it was to classify encrypted packets as compressed. The model misclassified  21.5\% of encrypted packets as compressed, while it misclassified  58.5\% of compressed packets as encrypted. 

\subsection{Feed-forward Neural Network}
The feed forward neural net reached an accuracy asymptote at 54.1\% measured by 10-fold cross validation. Modifying network hyperparameters resulted in minimal performance improvement. Any increase in the number of first layer neurons resulted in worse performance, likely due to overfitting. Training for more epochs resulted in rapid over-fitting, with accuracy plummeting past 100 epochs. Using stochastic gradient descent as opposed to an Adam optimizer resulted in slightly worse performance, averaging at 53.5\% with similar parameters. It appears that a standard feed-forward network lacks the functionality to achieve significant results for this problem, barring the existence of an extremely unconventional network topology.

\subsection{Convolutional Neural Network}
The CNN performed significantly better than the other two methods. Instead of hitting a rapid asymptote, it was possible to train this classifier to achieve relatively consistent results. This model averaged a testing accuracy of 66.9\% measured by 10-fold cross validation. The model still overfit rather quickly, with a reduction in accuracy occurring past 85 epochs. Adding additional layers and filters consistently decreased the accuracy of this model, suggesting that additional data would be needed to improve classification accuracy. 

\section{Discussion}
Despite the CNN's higher accuracy than the other two models, its performance is still insufficient for real world use. With an overall accuracy rate below 70\%, device auditing using the classifier would require extensive manual examination of packets flagged as compressed, since we lack a gold standard by which to detect compressed packets in the first place. 

Nevertheless, we believe that this project is an important first attempt at a machine learning problem which we hope will receive continued attention.  Our results demonstrate the detectability of underlying differences in entropy between compressed and encrypted data; a simple k-NN model was able to tag packets as compressed or encrypted with about 60\% accuracy, indicating that differences in packet variability between compressed and encrypted data are more practical for classification purposes than previously assumed.

While these results are insufficient for the automated privacy evaluation of consumer devices, we believe that significant improvements could be made. Given the overfitting issues observed with both feed-forward neural network and CNN models, we expect that the collection of additional data would improve classification accuracy. We are planning to collect a much larger dataset using the described pipeline, as well as by hand-labeling packet captures of traffic on an academic network. We also aim to explore solutions to the problems presented by variable packet sizes in real-world networking, such as evaluating the feasibility of padding too-small packets as compared simply discarding them. 

Additional improvements may be possible through adjustments in the machine learning implementations employed in this work. Further optimization of hyperparameters, including network architecture, may provide accuracy benefits; additionally, this work represents a relatively small sample of the possible machine learning approaches to this problem. Experimentation with other types of neural networks, such as long short-term memory networks (LSTMs) and autoencoders, may yield greater accuracy. We are continuing to work on model improvements, and we hope others in the community will take on this problem as well. 

\section{Conclusion}
The availability of a tool to automatically detect unencrypted packets regardless of encoding would be a major benefit for researchers and consumer protection advocates, allowing rapid evaluation of new networked devices for the privacy risks associated with the transmission of cleartext data. 
In this work, we examined the problem of distinguishing encrypted and compressed unencrypted network communications on a packet-by-packet basis. 
We trained three machine learning models to perform this task using only packet contents. 
A k-nearest neighbors model achieved an accuracy of 60\% while a 1D convolutional neural network achieved an accuracy of 66.9\%. 
These results, significantly higher than the 50\% expected accuracy of random guessing, demonstrate that the task is feasible and represent a baseline for further data collection and model improvements that will be required to achieve a useable accuracy for real world device auditing. 
We have made our dataset of compressed and encrypted packet contents publicly available and hope that others in the community will use this data to continue research on this problem.

\section*{Acknowledgments}
Thanks to the developers of SciKit Learn, numpy, dpkt, Keras, TensorFlow, Python, AES, Blowfish, Gzip, Bzip2, rar, 7zip, and RawCap for making their software freely available.  This work is supported by a Google Faculty Research Award, the National Science Foundation through awards CNS-1535796 and CNS-1539902, and the Princeton University Center for Information Technology Policy Internet of Things Consortium.

\bibliographystyle{ACM-Reference-Format}
\bibliography{CompressedCleartextDetection}

\end{document}